\documentclass [12pt]{article}
\usepackage{amsmath,amssymb,cite}

\usepackage{tabularx}
\usepackage[english]{babel}
\usepackage[dvips]{graphicx}
\usepackage{indentfirst}
\usepackage{epsfig}
\begin{document}

\title{The fate of the Wilson-Fisher fixed point in non-commutative $\phi^4$}

\author{Badis Ydri, Adel Bouchareb\\
}

\author{Badis Ydri\footnote{Email:ydri@stp.dias.ie,~badis.ydri@univ-annaba.org.}, Adel Bouchareb\footnote{Email:adel.bouchareb@univ-annaba.org.}\\
%Institut f\"{u}r Physik, Humboldt-Universit\"{a}t  zu Berlin, D-12489 Berlin-Germany}
%$^{+}$School of Theoretical Physics, Dublin Institute for Advanced Studies\\
% Dublin, Ireland.\\
Institute of Physics, BM Annaba University,\\
BP 12, 23000, Annaba, Algeria.
}

\maketitle
\abstract{In this article we study non-commutative vector sigma model with the most general $\phi^4$ interaction on Moyal-Weyl spaces. We compute the $2-$ and $4-$point functions to all orders in the large $N$ limit and then apply the approximate Wilson renormalization group recursion formula to study the renormalized coupling constants of the theory. The non-commutative  Wilson-Fisher fixed point interpolates between the commutative Wilson-Fisher fixed point of the Ising universality class which is found to lie at zero value of the critical coupling constant $a_*$ of the zero dimensional reduction of the theory, and a novel strongly interacting fixed point which lies at  infinite value of $a_*$ corresponding to maximal non-commutativity beyond which the two-sheeted structure of $a_*$ as a function of the dilation parameter disappears.}

%\tableofcontents

\section{Introduction}
A non-commutative scalar field theory is a non-local field theory in which the ordinary local point-wise multiplication of fields is replaced by
a non-local star product such as the Moyal-Weyl star product \cite{Groenewold:1946kp,Moyal:1949sk}. We suggest \cite{Alexanian:2000uz} and references therein for an elementary and illuminating discussion of the Moyal-Weyl product and other star products and their relation to the Weyl map \cite{weyl:1931}, coherent states \cite{Man'ko:1996xv,Perelomov:1986tf,klauder:1985}, Berezin  quantization \cite{Berezin:1974du} and deformation quantization \cite{Kontsevich:1997vb}.

The first study of the quantum theory of a non-commutative $\phi^4$ is found in \cite{Filk:1996dm} where it is shown that planar diagrams in the non-commutative theory are essentially identical to the planar diagrams in the commutative theory. More interestingly, it was found in \cite{Minwalla:1999px} that the renormalized one-loop action of a non-commutative $\phi^4$ suffers from an infrared divergence which is obtained when we send either the external momentum or the non-commutativity to zero. This non-analyticity at small momenta or small non-commutativity (IR) which is due to the high energy modes (UV) in virtual loops is termed the UV-IR mixing and it is intimately related to the structure of the non-planar diagrams of the theory. As it turns out this effect is expecteded in a non-local theory such as a non-commutative $\phi^4$.

Renormalization of the  non-commutative $\phi^4$ was studied for example in \cite{Chepelev:1999tt,Chepelev:2000hm,Grosse:2003aj,Grosse:2004yu,Grosse:2003nw,Rivasseau:2005bh,Gurau:2005gd}. The main observation of \cite{Grosse:2003aj} is that we can control the UV-IR mixing found in  non-commutative $\phi^4$ by modifying the large distance behavior of the free propagator through adding a harmonic oscillator potential to the kinetic term.  More precisely the UV-IR mixing of the theory is implemented precisely in terms of a certain duality symmetry of the new action which connects momenta and positions \cite{Langmann:2002cc}. The corresponding Wilson-Polchinski renormalization group equation \cite{Polchinski:1983gv,Keller:1990ej} of the theory can then be solved in terms of ribbon graphs drawn on Riemann surfaces. 

The existence of a regular solution of the Wilson-Polchinski equation together with the fact that we can scale to zero the coefficient of the harmonic oscillator potential in two dimensions leads to the conclusion that the standard non-commutative $\phi^4$ in two dimensions is renormalizable  \cite{Grosse:2003nw}. In four dimensions, the harmonic oscillator term seems to be essential for the renormalizability of the theory \cite{Grosse:2004yu}.

There are many other approaches to renormalization of quantum non-commutative $\phi^4$. See for example \cite{Becchi:2002kj,Becchi:2003dg,Gurau:2009ni,Griguolo:2001ez,Gurau:2008vd} and \cite{Sfondrini:2010zm}.

A very remarkable property of quantum non-commutative $\phi^4$ is the appearance of a new order in the theory termed the striped phase which was first computed in a one-loop self-consistent Hartree-Fock approximation in the seminal paper \cite{Gubser:2000cd}. For alternative derivations of this order see for example \cite{Chen:2001an,Castorina:2003zv}. It is believed that the perturbative UV-IR mixing is only a manifestation of this more profound property. As it turns out, this order should be called more appropriately a non-uniform ordered phase in contrast with the usual uniform ordered phase of the Ising universality class and it is related to spontaneous breaking of translational invariance. It was numerically  observed in $d=4$ in \cite{Ambjorn:2002nj} and in $d=3$ in \cite{Bietenholz:2004xs} where the Moyal-Weyl space was approximated by a non-commutative fuzzy torus \cite{Ambjorn:2000cs}. The beautiful result of \cite{Bietenholz:2004xs} shows explicitly that the minimum of the model shifts to a non-zero value of the momentum indicating a non-trivial condensation and hence spontaneous breaking of translational invariance.  

In summary the phase diagram of quantum non-commutative $\phi^4$ consists of three phases. The usual disordered and uniform ordered phases together with the non-uniform (striped) ordered phase and thus a triple point must exist. In \cite{Gubser:2000cd} and also in \cite{Chen:2001an}, it is conjectured that this point is a Lifshitz point which is a multi-critical point at which a disordered, a homogeneous (uniform) ordered and a spatially modulated (non-uniform) 
ordered phases meet \cite{Hornreich:1975zz}. We note that reference  \cite{Chen:2001an} uses Wilson renormalization group approach \cite{Wilson:1973jj} to derive the Wilson-Fisher fixed point of the theory at one-loop which is found to suffer from an instability at large non-commutativity.
% to this problem.

The non-commutative fuzzy torus used to regularized  non-commutative $\phi^4$ is a matrix model which can be mapped to a lattice. Another matrix regularization of  non-commutative $\phi^4$ can be found  in \cite{Langmann:2003if} and also in \cite{Steinacker:2005wj} which emphasizes connection to fuzzy spaces \cite{Balachandran:2005ew,O'Connor:2003aj}. The phase structure of non-commutative $\phi^4$ in $d=2$ and $d=3$ using the fuzzy sphere \cite{Hoppe:1982,Madore:1991bw} regularization was studied extensively in \cite{Martin:2004un,GarciaFlores:2009hf,Panero:2006bx,Medina:2007nv,Das:2007gm}. Again the phase diagram consists of three phases: a disordered phase, a uniform ordered phases and a non-uniform ordered phase which meet at a triple point. In this case it is well established that the transitions from the disordered phase to the non-uniform ordered phase and from the non-uniform ordered phase to the uniform ordered phase originate from the one-cut/two-cut transition in the quartic hermitian matrix model \cite{Brezin:1977sv,Shimamune:1981qf}. See also \cite{DiFrancesco:1993nw,eynard}. This was also confirmed analytically by the multi-trace approach  of \cite{O'Connor:2007ea,Saemann:2010bw} which relies on the expansion of the kinetic term in the action instead of the usual expansion in the interaction which is very reminiscent  to the hopping parameter expansion on the lattice \cite{Montvay:1994cy,Smit:2002ug}. 

The non-uniform ordered phase on the fuzzy sphere (sometimes also called the matrix phase) goes to the striped phase on the Moyal-Weyl plane in appropriate flattening limit. Finally there is a strong evidence that the non-uniform ordered phase should be present on all non-commutative spaces regardless of the dimension.

In this article we will attempt to understand the phase structure of non-commutative $\phi^4$ near the Wilson-Fisher fixed point, i.e. the transition disordered/uniform ordered and how this critical behavior changes with the non-commutativity until it merges with the transition disordered/non-uniform ordered. We will consider a large vector $O(N)$ sigma model where all leading Feynman diagrams can be taken into consideration and employ the approximate Wilson renormalization group equation to study the renormalized action of the theory. 

This article is organized as follows. In section $2$ we will introduce the non-commutative vector sigma model with the most general $\phi^4$ interaction on the Moyal-Weyl spaces ${\bf R}_{\theta}^d$, and then we will compute in the large $N$ limit the $2-$ and $4-$point functions to all orders. In section $3$ we  will apply the approximate Wilson renormalization group recursion formula to study the renormalized coupling constants of the theory. In particular we will derive the renormalization group equations and then calculate the fixed points of the theory. In section $4$ we give our conclusion and outlook.

%\tableofcontents
\section{The Non-Commutative $O(N)$ Sigma Model}
\paragraph{The Cumulant Expansion:}
We will consider in this note the $\phi^4$ action
\begin{eqnarray}
S=\int d^dx \Phi_a(-{\partial}_i^2+{\mu}^2)\Phi_a+S_{\rm int}.
\end{eqnarray}
\begin{eqnarray}
S_{\rm int}=u\int d^dx~(\Phi_a*\Phi_a)^2+v\int d^dx~(\Phi_a*\Phi_b)^2.
\end{eqnarray}
We recall the star product in the form (with $\int_p=\int d^dp/(2\pi)^d$)
\begin{eqnarray}
f*g(x)=\int_p \int_k~\tilde{f}(p)\tilde{g}(k)~e^{i(p+k)x}~e^{-\frac{i}{2}p\wedge k}~,~p\wedge k=\theta_{ij}p_ik_j.
\end{eqnarray}
The propagator and the vertex in momentum space read 
\begin{eqnarray}
\Delta_{ab}(x-x^{'})=\int_p \Delta_{ab}(p)~e^{ip(x-x^{'})}~,~{\Delta}_{ab}(p)=\frac{1}{2}\frac{\delta_{ab}}{p^2+\mu^2}.
\end{eqnarray}
\begin{eqnarray}
S_{\rm int}&=&\int_{k_1}...\int_{k_4}\tilde{f}_a(k_1)\tilde{f}_a(k_2)\tilde{f}_b(k_3)\tilde{f}_b(k_4)~(2\pi)^d\delta^d(k_1+...+k_4)\tilde{V}(k_1,k_2,k_3,k_4).
\end{eqnarray}
\begin{eqnarray}
\tilde{V}(k_1,k_2,k_3,k_4)=\frac{u_0}{N}\cos\frac{k_1\wedge k_2}{2}\cos\frac{k_3\wedge k_4}{2}+\frac{v_0}{2N}(e^{\frac{i}{2}(k_1\wedge k_3+k_2\wedge k_4)}+e^{\frac{i}{2}(k_1\wedge k_4+k_2\wedge k_3)}).
\end{eqnarray}
We decompose the fields $\Phi_a(x)$ into backgrounds $\phi_a(x)$ which contain slow modes, i.e. modes with momenta less or equal than $\rho\Lambda$ and fluctuations $f_a(x)$ which contain fast modes, i.e. modes with momenta  larger  than $\rho\Lambda$ where $0<\rho<1$. We are interested in the partition function 
\begin{eqnarray}
Z=\int d\Phi_a~e^{-S[\Phi]}&=&\int d\phi_a~e^{-S[\phi]}\int df_a~e^{-S[f]}e^{-\sigma(\phi,f)}\nonumber\\
&=&\int d\phi_a~e^{-S[\phi]}<e^{-\sigma(\phi,f)}>\int df_a e^{-S[f]}.
\end{eqnarray}
We have defined
\begin{eqnarray}
<{\cal O}>=\frac{\int df_a~{\cal O}~e^{-S[f]}}{\int df_a e^{-S[f]}}.
\end{eqnarray}
The action $\sigma(\phi,f)$ is of the form $\sigma(\phi,f)={\cal M}_1+{\cal M}_2+{\cal M}_3$ where ${\cal M}_n$ contains all the terms which are of order $n$ in the background field $\phi_a$. Clearly the effective action is given by $S_{\rm eff}[\phi]=S[\phi]+\Delta S_{\rm eff}[\phi]$ where $\Delta S_{\rm eff}[\phi]$ is defined by 
\begin{eqnarray}
<e^{-\sigma(\phi,f)}>&=&e^{-\Delta S_{\rm eff}[\phi]}.
\end{eqnarray}
By using the symmetry under $\phi_a\longrightarrow -\phi_a$ and momentum conservation, we compute up to the $4$th order in the slow fields $\phi_a$ the non perturbative expansion 

\begin{eqnarray}
\Delta S_{\rm eff}[\phi]&=&<{\cal M}_2>_{\rm co}-\frac{1}{2}<{\cal M}_1^2>_{\rm co}-\frac{1}{2}<{\cal M}_2^2>_{\rm co}+\frac{1}{2}<{\cal M}_1^2{\cal M}_2>_{\rm co}-\frac{1}{24}<{\cal M}_1^4>_{\rm co}.\label{eff}\nonumber\\
\end{eqnarray}
At this stage we employ the large $N$ limit. After appropriate rescaling, the propagator comes with $1/N$ factor, the vertex comes with a factor of $N$ and the contraction of a vector index yields a factor of $N$. There exists a non-trivial $1/N$ expansion only if $u,v\longrightarrow 0$ when $N\longrightarrow \infty$ such that $u_0=u N$ and $v_0=v N$ is kept fixed. By inspection it is found that all terms %($O(N)$-singlets) 
of the form  $..*\phi_a*..*f_a*..$ are subleading in the large $N$ limit. In other words we can set in the large $N$ limit ${\cal M}_1,{\cal M}_3\longrightarrow 0$ and 
\begin{eqnarray}
{\cal M}_2&=&2u\int d^dx~\phi_a*\phi_a*f_b*f_b+2v\int d^dx~\phi_a*f_b*\phi_a*f_b.
\end{eqnarray}
As a result the final form of the effective action is given explicitly by the very simple cumulant expansion 
\begin{eqnarray}
\Delta S_{\rm eff}[\phi]&=&<{\cal M}_2>_{\rm co}-\frac{1}{2}<{\cal M}_2^2>_{\rm co}.
\end{eqnarray}
\paragraph{The Renormalized Action:}
Next we will give the exact solution of the model in the large $N$ limit by computing formally all Feynman diagrams contributing to the $2-$ and the $4-$point functions.
%\subsection{Mass and Wave Function Renormalization}

\begin{figure}[htbp]
\begin{center}
\includegraphics[width=4cm,angle=0]{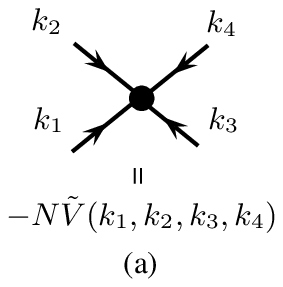}
\includegraphics[width=10.0cm,angle=0]{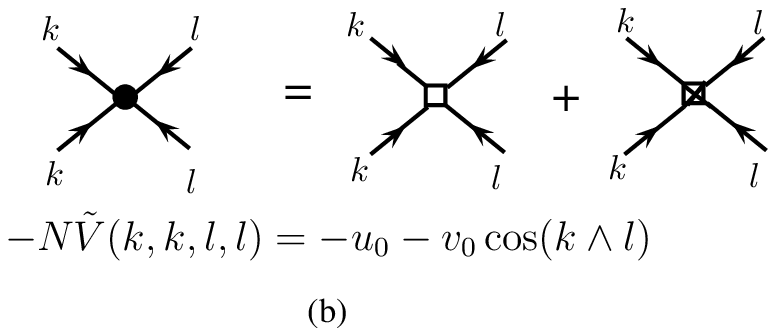}
\caption{The vertex of non-commutative $O(N)$ sigma model.}\label{figure1}
\end{center}
\end{figure}

 We start with the correction to the quadratic action given by
\begin{eqnarray}
\Delta S_{\rm quad}[\phi]&=&<{\cal M}_2>_{\rm co}.
\end{eqnarray}
Explicitly we have
\begin{eqnarray}
\Delta S_{\rm quad}[\phi]=2u\int d^dx~<\phi_a*\phi_a*f_b*f_b>_{\rm co}+2v\int d^dx~<\phi_a*f_b*\phi_a*f_b>_{\rm co}.
\end{eqnarray}
After a somewhat lengthy calculation we reach the result (we show terms up to three loops)
\begin{eqnarray}
\Delta S_{\rm quad}[\phi]=\int_{p_1}\tilde{\phi}_a(p_1)\Delta \Gamma_2(p_1)\tilde{\phi}_a(-p_1)
\end{eqnarray}
\begin{eqnarray}
\Delta \Gamma_2(p_1)=u_0\int_{k_1}\Delta \Gamma_2(p_1,k_1)+v_0\int_{k_1}\Delta \Gamma_2(p_1,k_1)\cos k_1\wedge p_1.%e^{-ip_1\wedge k_1}.
\end{eqnarray}
\begin{eqnarray}
\Delta \Gamma_2(p_1,k_1)&=&\Delta(k_1)-\Delta^2(k_1)\int_l \Delta(l) (u_0+v_0\cos k_1\wedge l)+\Delta^2(k_1)\int_l\Delta^2(l)(u_0+v_0\cos k_1\wedge l)\nonumber\\
&\times &\int_{l^{'}}\Delta (l^{'})(u_0+v_0\cos l \wedge l^{'})+ \Delta^3(k_1)\bigg(\int_l\Delta(l)(u_0+v_0\cos k_1\wedge l)\bigg)^2+...
\end{eqnarray}
We recognize this series as a sum of all bubble graphs with a combination of the planar vertex $-u_0$ and the non planar vertex $-v_0\cos p\wedge k$ where $p$ and $k$ are the momenta flowing into the vertex. Therefore in the large $N$ limit the effective Feynman vertices for the fields $f$ are simply those shown on graph $(b)$ of figure (\ref{figure1}). The full vertex is shown on graph $(a)$ of figure (\ref{figure1}). 

\begin{figure}[htbp]
\begin{center}
\includegraphics[width=12.0cm,angle=0]{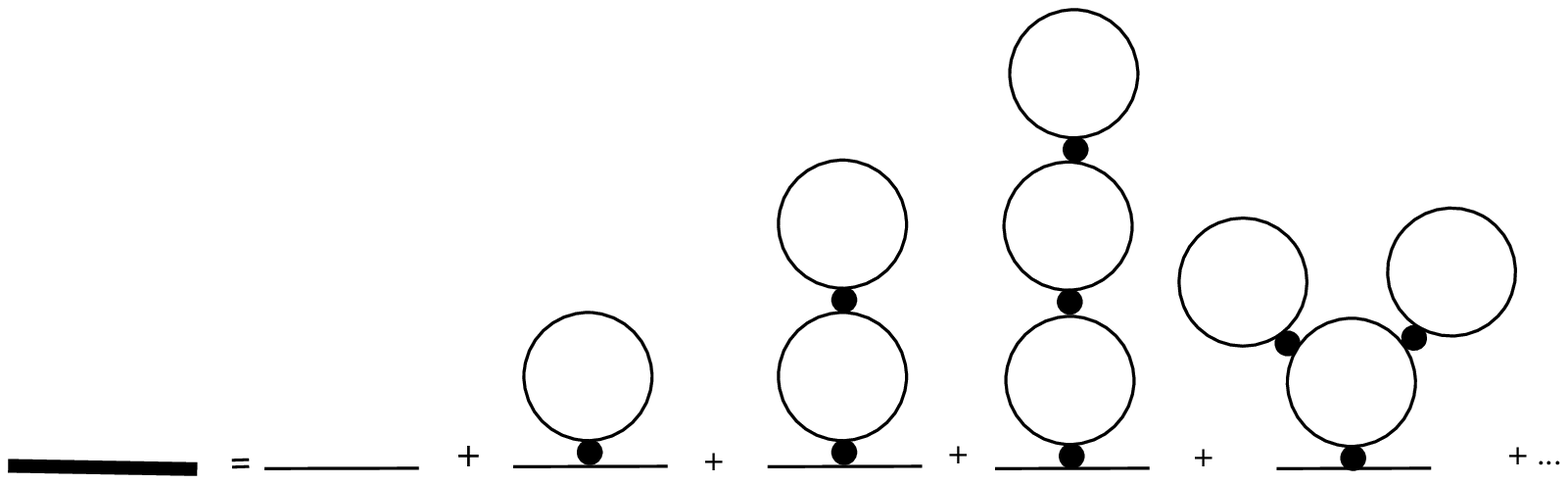}
\caption{The dressed propagator of non-commutative $O(N)$ sigma model.}\label{figure2}
\end{center}
\end{figure}

The three-loop diagrams shown in $\Delta \Gamma_2(p_1,k_1)$ can be represented by the Feynman diagrams on figure (\ref{figure2}). It is not difficult to convince ourselves that in the large $N$ limit all leading higher loops correspond to bubble diagrams. The series can be summed up. Indeed we can easily verify that
\begin{eqnarray}
\Delta \Gamma_2(p_1)&=&\delta \mu^2_P+\delta \mu^2_{NP}(p_1).
\end{eqnarray}
\begin{eqnarray}
\delta \mu^2_P&=&u_0\int_{k_1}\frac{1}{k_1^2+\mu^2+\delta \mu^2_P+\delta \mu^2_{NP}(k_1)}.
\end{eqnarray}
\begin{eqnarray}
\delta \mu^2_{NP}(p_1)&=&v_0\int_{k_1}\frac{1}{k_1^2+\mu^2+\delta \mu^2_P+\delta \mu^2_{NP}(k_1)}\cos k_1\wedge p_1.
\end{eqnarray}
Firstly, we remark that the $2-$point proper vertex obtained in \cite{Gubser:2000cd} using the Hartree-Fock approximation in the $O(1)$ non-commutative $\phi^4$ is similar in structure to our result here. Secondly, as opposed to the commutative theory, mass renormalization here depends on the external momentum. This means in particular that there will be, in general, a wave function renormalization in the noncommutative large $N$ linear sigma model.

\begin{figure}[htbp]
\begin{center}
\includegraphics[width=14.0cm,angle=0]{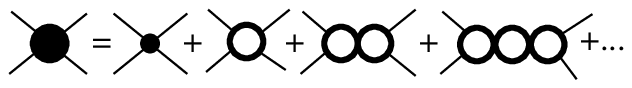}
\includegraphics[width=10.0cm,angle=0]{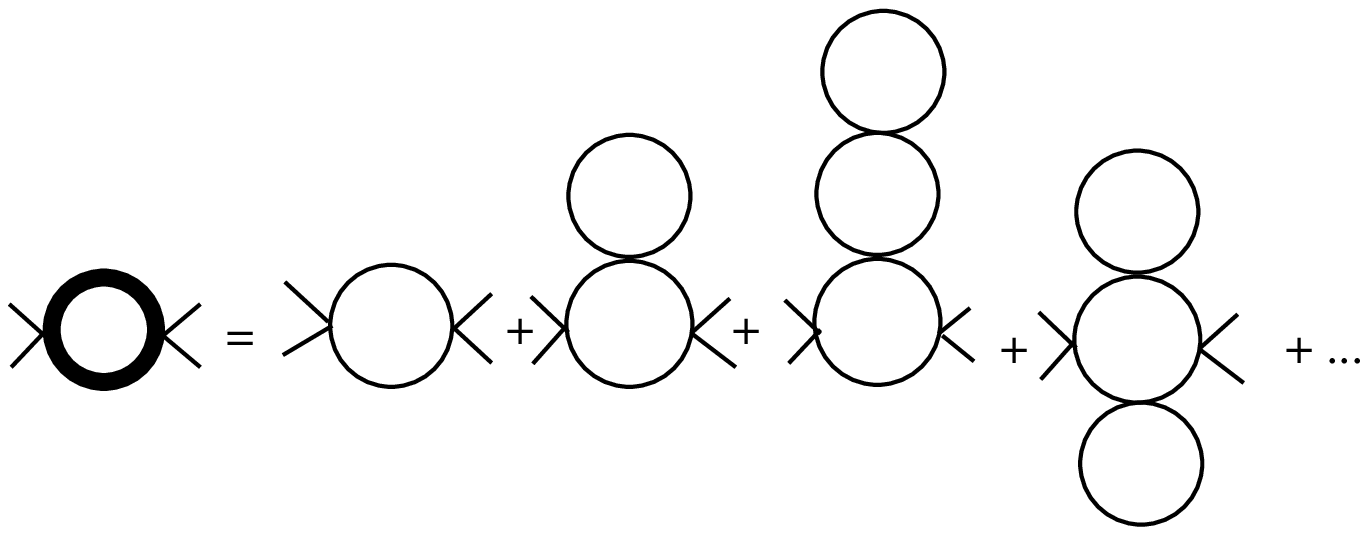}
\includegraphics[width=12.0cm,angle=0]{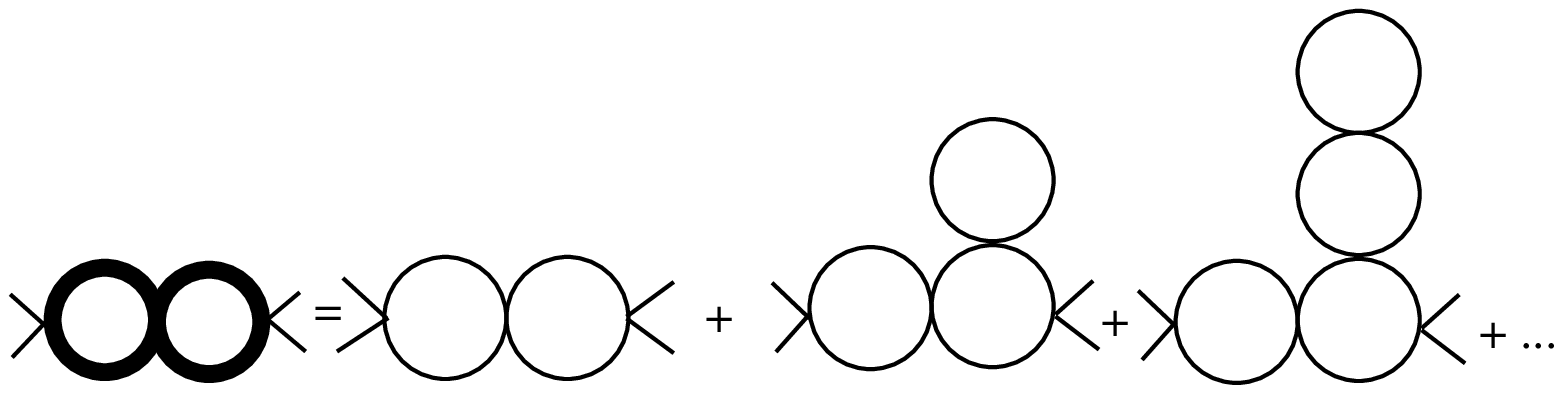}
\includegraphics[width=8.0cm,angle=0]{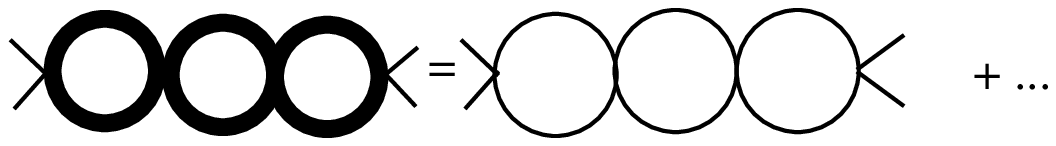}
\caption{The dressed vertex of non-commutative $O(N)$ sigma model.}\label{figure3}
\end{center}
\end{figure}

%\subsection{Interaction Renormalization}
Now we discuss the $4-$point function. The full correction to the quartic action in the large $N$ limit is given by
\begin{eqnarray}
\Delta S_{\rm int}[\phi]&=&-\frac{1}{2}<{\cal M}_2^2>_{\rm co}.
\end{eqnarray}
Explicitly we have
\begin{eqnarray}
\Delta S_{\rm int}[\phi]&=&-2\int_{p_1}..\int_{p_4}\tilde{\phi}_a(p_1)\tilde{\phi}_a(p_2)\tilde{\phi}_b(p_3)\tilde{\phi}_b(p_4)~\int_{k_1}..\int_{k_4}\tilde{V}(p_1,p_2,k_1,k_2)~\tilde{V}(p_3,p_4,k_3,k_4)\nonumber\\
%&\times &(2\pi)^d\delta^d(p_1+p_2+k_1+k_2)(2\pi)^d\delta^d(p_3+p_4+k_3+k_4)<\tilde{f}_c(k_1)\tilde{f}_c(k_2)\tilde{f}_d(k_3)\tilde{f}_d(k_4)>_{\rm co}.\nonumber\\
&\times &<\tilde{f}_c(k_1)\tilde{f}_c(k_2)\tilde{f}_d(k_3)\tilde{f}_d(k_4)>_{\rm co}.
\end{eqnarray}
In the above equation we have adopted a new convention, namely $\tilde{V}(k_1,k_2,k_3,k_4)$ contains now a delta function $(2\pi)^d\delta^d(k_1+k_2+k_3+k_4)$. We show after another much longer calculation that the leading Feynman diagrams contributing to the  correction of the quartic interaction in the large $N$ limit are only those given on figure (\ref{figure3}).  These diagrams can be partially summed up to give the relatively compact result 
%(with the convention that  $\tilde{V}(k_1,k_2,k_3,k_4)$ contains now a delta function $(2\pi)^d\delta^d(k_1+k_2+k_3+k_4)$)
\begin{eqnarray}
\Delta S_{\rm int}&=&-\int_{p_1}...\int_{p_4}\tilde{\phi}_a(p_1)\tilde{\phi}_a(p_2)\tilde{\phi}_b(p_3)\tilde{\phi}_b(p_4)~\Delta \tilde{V}(p_1,p_2,p_3,p_4).
\end{eqnarray}
\begin{eqnarray}
\Delta \tilde{V}(p_1,p_2,p_3,p_4)&=&N\int_{k_1}\int_{k_2}\tilde{V}(p_1,p_2,k_1,k_2)\tilde{V}(p_3,p_4,-k_1,-k_2)~\Delta_E(k_1)\Delta_E(k_2)\nonumber\\
&-&N^2\int_{k_1}\int_{k_2}\int_{k_3}\int_{k_4}\tilde{V}(p_1,p_2,k_1,k_2)\tilde{V}(p_3,p_4,k_3,k_4)~\Delta_E(k_1)..\Delta_E(k_4)\nonumber\\
&\times &\tilde{V}(-k_1,-k_2,-k_3,-k_4)\nonumber\\
&+&N^3\int_{k_1}\int_{k_2}\int_{k_3}\int_{k_4}\tilde{V}(p_1,p_2,k_1,k_2)\tilde{V}(p_3,p_4,k_3,k_4)~\Delta_E(k_1)..\Delta_E(k_4)\nonumber\\
&\times &\int_{l_1}\int_{l_2}\tilde{V}(-k_1,-k_2,l_1,l_2)\tilde{V}(-k_3,-k_4,-l_1,-l_2)~\Delta_E(l_1)\Delta_E(l_2)\nonumber\\
%&-&N^4\int_{k_1}\int_{k_2}\int_{k_3}\int_{k_4}\tilde{V}(p_1,p_2,k_1,k_2)\tilde{V}(p_3,p_4,k_3,k_4)~\Delta_E(k_1)\Delta_E(k_2)\Delta_E(k_3)\Delta_E(k_4)\nonumber\\
%&\times &\int_{l_1}\int_{l_2}\int_{l_3}\int_{l_4}\tilde{V}(-k_1,-k_2,l_1,l_2)\tilde{V}(-k_3,-k_4,l_3,l_4)~\Delta_E(l_1)\Delta_E(l_2)\Delta_E(l_3)\Delta_E(l_4)\nonumber\\
%&\times &\tilde{V}(-l_1,-l_2,-l_3,-l_4)\nonumber\\
&+&....
\end{eqnarray}
The terms proportional to $N^k$ correspond to Feynman graphs with $k$ loops. For example the terms proportional to $N$, $N^2$, $N^3$  in the above equation are represented by the dressed one-loop, two-loop and three-loop Feynman diagrams shown on figure (\ref{figure3}). The exact propagator which we will denote in Feynman diagrams by a solid line is obviously defined by
\begin{eqnarray}
\Delta_E(k)&=&\frac{1}{k^2+\mu^2+\Delta \Gamma_2(k)}.
\end{eqnarray}

\section{Renormalization Group Equations and Fixed Points}
\paragraph{Wilson Contraction:}
%\subsection{The Renormalized Action}
In the previous section, we have solved the theory in the large $N$ limit by computing the $2-$ and $4-$point functions to all orders. However, we still do not really know how to evaluate explicitly the contributing Feynman diagrams. In this section we will provide an approximation scheme which will give us a very good insight into the structure of the theory.

More precisely, we will apply in this section the renormalization group program of Wilson \cite{Wilson:1973jj} as applied (with great success) in \cite{Ferretti:1995zn,Nishigaki:1996ts} to ordinary vector models  and to hermitian matrix models in the large $N$ limit. Their method can be summarized as follows:
\begin{itemize}
\item[$1)$]We will split the field into a background and a fluctuation and then integrate the fluctuation obtaining therefore an effective action for the background field alone. This has already been done in the section $2$.
\item[$2)$]We will keep, following Wilson, only induced corrections to the terms that are already present in the classical action. This also has already been in section $2$ by calculating only the $2-$ and $4-$point functions. 
%the In some sense this is equivalent to taking the limit $\theta\Lambda^2\longrightarrow 0$.
\item[$3)$]We perform the so-called Wilson contraction which consists in estimating momentum loop integrals using the following three approximations or rules:
\begin{itemize}
\item{}{\bf Rule} $1$: All external momenta which are wedged with internal momenta will be set to zero.
\item{}{\bf Rule} $2$: We approximate every internal propagator $\Delta({k})$ by $\Delta({\lambda})$ where ${\lambda}$ is a typical momentum in the range $\rho\Lambda\leq\lambda\leq \Lambda$. 
\item{}{\bf Rule} $3$: We replace every internal momentum loop integral $\int_{{k}}$ by a typical volume.
 \end{itemize}
  The two last approximations are equivalent to the reduction of all loop integrals to their zero dimensional counterparts. We will choose for simplicity $\lambda=\Lambda$ and the typical volume is  
\begin{eqnarray}
\int_{{k}}1=\frac{\Lambda^d}{(2\pi)^d}\frac{2\pi^{\frac{d}{2}}}{\Gamma(\frac{d}{2})} \frac{1}{d}[1-\rho^d]=\Lambda^dV_d.\label{vd}
\end{eqnarray}
These two approximations are quite natural in the limit $\rho\longrightarrow 1$. 

 As it turns out we do not need to use the first approximation in estimating the $2-$point function. In fact rule $1$ was proposed first in the context of a non-commutative  $\phi^4$ theory in \cite{Chen:2001an} in order to simply the calculation of the $4-$point function. In some sense the first approximation is equivalent to taking the limit $\bar{\theta}=\theta\Lambda^2\longrightarrow 0$.

\item[$4)$]The last step in the renormalization group program of Wilson consists in rescaling the momenta so that the cutoff is restored to its original value. We can then obtain renormalization group recursion equations which relate the new values of the coupling constants to the old values.
\end{itemize}  
This strategy was applied to the non-commutative $\phi^4$ in \cite{Chen:2001an}. The difference with our current case lies in the central fact that in the context of a large $O(N)$ sigma model we can take into account all leading diagrams and not only the one-loop diagrams and thus the result is non-perturbative.

We start by writing the full action: the classical+the complete quantum corrections in the large $N$ limit for the quadratic and quartic terms as

\begin{eqnarray}
S_{\rm eff}&=&\int_{p\leq \rho\Lambda}\tilde{\phi}_a(p)\bigg(p^2+\mu^2+\Delta \Gamma_2(p)\bigg)\tilde{\phi}_a(-p)\nonumber\\
&+&\int_{p_1\leq \rho \Lambda}...\int_{p_4\leq \rho\Lambda}\tilde{\phi}_a(p_1)\tilde{\phi}_a(p_2)\tilde{\phi}_b(p_3)\tilde{\phi}_b(p_4)~(2\pi)^d\delta^d(p_1+...+p_4)~\tilde{V}_{\rm eff}(p_1,p_2,p_3,p_4).\nonumber\\
\end{eqnarray}

\paragraph{The Quadratic Action:}

Recall that
\begin{eqnarray}
\Delta \Gamma_2(p)=\int_{\rho\Lambda\leq k\leq \Lambda}\frac{1}{k^2+\mu^2+\Delta\Gamma_2(k)}~\bigg(u_0+v_0\cos k\wedge p\bigg).
\end{eqnarray}
We will employ the dimensionless quantities: $\bar{p}=p/\Lambda$, $\bar{k}=k/\Lambda$, $\bar{u}_0=\Lambda^{d-4}u_0$, $\bar{v}_0=\Lambda^{d-4}v_0$, $\bar{\mu}^2=\mu^2/\Lambda^2$ and $\Delta \bar{\Gamma}_2(\bar{p})=\Delta\Gamma_2(\Lambda \bar{p})/\Lambda^2$. By doing the integral over the $d-1$ angles involved in $d^dk$ we arrive at the result
\begin{eqnarray}
\Delta \bar{\Gamma}_2(\bar{p})=\frac{1}{(2\pi)^d}\frac{2\pi^{\frac{d}{2}}}{\Gamma(\frac{d}{2})} \int_{\rho \leq \bar{k}\leq 1}\bar{k}^{d-1} d\bar{k} \frac{1}{\bar{k}^2+\bar{\mu}^2+\Delta\bar{\Gamma}_2(\bar{k})}~\bigg(\bar{u}_0+\bar{v}_0\tilde{X}_{d-1}(\bar{\theta} \bar{k}\bar{p})\bigg).
\end{eqnarray}
\begin{eqnarray}
\tilde{X}_{d-1}(x)&=&\bigg(\frac{2}{x}\bigg)^{\frac{d}{2}-1}\Gamma\bigg(\frac{d}{2}\bigg)J_{\frac{d-2}{2}}(x).
\end{eqnarray}
The explicit solution is
\begin{eqnarray}
\Delta \bar{\Gamma}_2(\bar{p})&=&\Delta \bar{\Gamma}_2^0(\bar{p})+\Delta \bar{\Gamma}_2^1(\bar{p})+\Delta \bar{\Gamma}_2^2(\bar{p})+...%\Delta \bar{\Gamma}_2^3(\bar{p})+...
\end{eqnarray}
The term $\Delta \bar{\Gamma}_2^n(\bar{p})$ corresponds to the sum of $n-$loop Feynman diagrams contributing to the $2-$point function. By applying rules $2$ and $3$ of Wilson contraction we get the approximations
\begin{eqnarray}
\Delta \bar{\Gamma}_2^0(\bar{\lambda})\longrightarrow V_d \Delta(\bar{\lambda})~\bigg(\bar{u}_0+\bar{v}_0\tilde{X}_{d-1}(\bar{\theta} \bar{\lambda}^2)\bigg),
\end{eqnarray}
\begin{eqnarray}
\Delta \bar{\Gamma}_2^1(\bar{\lambda})\longrightarrow -V_d^2 \Delta^3(\bar{\lambda})~\bigg(\bar{u}_0+\bar{v}_0\tilde{X}_{d-1}(\bar{\theta} \bar{\lambda}^2)\bigg)^2,
\end{eqnarray}
\begin{eqnarray}
\Delta \bar{\Gamma}_2^2(\bar{\lambda})\longrightarrow 2V_d^3 \Delta^5(\bar{\lambda})~\bigg(\bar{u}_0+\bar{v}_0\tilde{X}_{d-1}(\bar{\theta} \bar{\lambda}^2)\bigg)^3,
\end{eqnarray}
%\begin{eqnarray}
%\Delta \bar{\Gamma}_2^3(\bar{p})\longrightarrow -5V_d^4\Delta^7(\bar{\lambda})~\bigg(\bar{u}_0+\bar{v}_0\tilde{X}_{d-1}(\bar{\theta} \bar{\lambda}^2)\bigg)^4,
%\end{eqnarray}
etc. Hence we obtain
\begin{eqnarray}
\Delta \bar{\Gamma}_2(\bar{p})&=&V_d\Delta(\bar{\lambda})~\bigg(\bar{u}_0+\bar{v}_0\tilde{X}_{d-1}(\bar{\theta} \bar{\lambda}\bar{p})\bigg)\bigg[1-V_d\Delta^2(\bar{\lambda})~\bigg(\bar{u}_0+\bar{v}_0\tilde{X}_{d-1}(\bar{\theta} \bar{\lambda}^2)\bigg)\nonumber\\
&+&2V_d^2\Delta^4(\bar{\lambda})~\bigg(\bar{u}_0+\bar{v}_0\tilde{X}_{d-1}(\bar{\theta} \bar{\lambda}^2)\bigg)^2-5V_d^3\Delta^6(\bar{\lambda})~\bigg(\bar{u}_0+\bar{v}_0\tilde{X}_{d-1}(\bar{\theta} \bar{\lambda}^2)\bigg)^3\nonumber\\
&+&...\bigg].
\end{eqnarray}
The expansion between brackets is recognized as the expansion of the connected $2-$point function of the zero-dimensional vector model which is given by the function \cite{Hikami:1978ya,Nishigaki:1996ts}
\begin{eqnarray}
C_2(a)=\frac{\sqrt{1+4a}-1}{2a}=1-a+2a^2-5a^3+...
\end{eqnarray}
The final result is therefore of the form
\begin{eqnarray}
\Delta \bar{\Gamma}_2(\bar{p})&=&V_d\Delta(\bar{\lambda})~\bigg(\bar{u}_0+\bar{v}_0\tilde{X}_{d-1}(\bar{\theta} \bar{\lambda}\bar{p})\bigg)C_2(a).
%\frac{\bar{u}_0+\bar{u}_0^{'}\tilde{X}_{d-1}(\bar{\theta} \bar{\lambda}\bar{p})}{\bar{u}_0+\bar{u}_0^{'}\tilde{X}_{d-1}(\bar{\theta} \bar{\lambda}^2)}(\bar{\lambda}^2+\bar{\mu}^2)aC_2(a).
%v_d(\rho)\Delta(\bar{\lambda})~\bigg(\bar{u}_0+\bar{u}_0^{'}\tilde{X}_{d-1}(\bar{\theta} \bar{\lambda}\bar{p})\bigg)C_2(a).
\end{eqnarray}
%\begin{eqnarray}
%\Delta \bar{\Gamma}_2(\bar{p})&=&v_d(\rho)\Delta(\bar{\lambda})~\bigg(\bar{u}_0+\bar{u}_0^{'}\tilde{X}_{d-1}(\bar{\theta} \bar{\lambda}\bar{p})\bigg)C_2(a).
%\end{eqnarray}
\begin{eqnarray}
a=V_d\Delta^2(\bar{\lambda})~\bigg(\bar{u}_0+\bar{v}_0\tilde{X}_{d-1}(\bar{\theta} \bar{\lambda}^2)\bigg).
\end{eqnarray}
Following  \cite{Ferretti:1995zn,Nishigaki:1996ts} we will choose the typical momentum $\bar{\lambda}=1$, and for simplicity we will also choose $\bar{u}_0=\bar{v}_0$. We get therefore
%In this case
\begin{eqnarray}
a=V_d\frac{\bar{u_0}}{(1+\bar{\mu}^2)^2}~\bigg(1+\tilde{X}_{d-1}(\bar{\theta})\bigg).\label{a}
\end{eqnarray}
%And
\begin{eqnarray}
\Delta \bar{\Gamma}_2(\bar{p})&=&\frac{\bar{u}_0}{1+\bar{\mu}^2}V_d~\bigg(1+\tilde{X}_{d-1}(\bar{\theta}\bar{p})\bigg)C_2(a).\label{wave}
\end{eqnarray}
%In obtaining this result we have not used rule $1$ of Wilson contraction. 
The first main observation is that non-commutative $O(N)$ sigma model involves a wave function renormalization as opposed to the commutative $O(N)$ sigma model. By expanding around small external momenta, we observe that this renormalization comes with the wrong (negative) sign which signals the existence of an instability in the theory \cite{Chen:2001an}. 

Secondly, the set of approximations (the so-called Wilson contraction) used in estimating the leading diagrams in the large $N$ limit consists in three rules where the first rule says that any external momentum which is wedge with an internal momentum must be set to zero. This rule was, in fact, not used in the estimation of the tadpole graphs contributing to the $2$-point function and as a consequence the calculation of the $2$-point function is, in a sense, more exact. 
%Using this rule in the $2$-point function would simply  imply that we must set $\delta Z=1$.
By applying rule $1$ of Wilson contraction we get the simpler answer 
 \begin{eqnarray}
\Delta \bar{\Gamma}_2(\bar{p})&=&2\frac{\bar{u}_0}{1+\bar{\mu}^2}V_d~C_2(a)\equiv \delta\bar{\mu}^2.
\end{eqnarray}
The quadratic part of the action becomes
%\begin{eqnarray}
%\int_{p\leq \rho\Lambda}\tilde{\phi}_a(p)\bigg(p^2+\mu^2+\Delta \Gamma_2(p)\bigg)\tilde{\phi}_a(-p)&=&\Lambda^{d+2}\int_{\bar{p}\leq \rho}\tilde{\phi}_a(\Lambda \bar{p})\bigg[\bigg(1-\delta Z\bigg)\bar{p}^2+\bigg(\bar{\mu}^2+\delta\bar{\mu}^2\bigg)+...\bigg]\tilde{\phi}_a(-\Lambda \bar{p}).\nonumber\\\label{quadraticS}
%\end{eqnarray}
\begin{eqnarray}
\int_{p\leq \rho\Lambda}\tilde{\phi}_a(p)\bigg(p^2+\mu^2+\Delta \Gamma_2(p)\bigg)\tilde{\phi}_a(-p)&=&\Lambda^{d+2}\int_{{p}\leq \rho}\tilde{\phi}_a(\Lambda {p})\bigg[{p}^2+\bigg(\bar{\mu}^2+\delta\bar{\mu}^2\bigg)\bigg]\tilde{\phi}_a(-\Lambda {p}).\nonumber\\
\label{quadraticS}
\end{eqnarray}
%\begin{eqnarray}
%\int_{p\leq \rho\Lambda}\tilde{\phi}_a(p)\bigg(p^2+\mu^2+\Delta \Gamma_2(p)\bigg)\tilde{\phi}_a(-p)&=&\Lambda^2\int_{\bar{p}\leq \rho}\tilde{\phi}_a(\Lambda \bar{p})\bigg[\bigg(1-\frac{\bar{\theta}^2(1+\bar{\mu}^2)aC_2(a)}{2d(1+\tilde{X}_{d-1}(\bar{\theta}))}\bigg)\bar{p}^2\nonumber\\
%&+&\bigg(\bar{\mu}^2+\frac{2(1+\bar{\mu}^2)aC_2(a)}{1+\tilde{X}_{d-1}(\bar{\theta})}\bigg)\bigg]\tilde{\phi}_a(-\Lambda \bar{p}).
%\end{eqnarray}
%This needs to be compared with the original quadratic action given by
%\begin{eqnarray}
%\int_{p\leq \Lambda}\tilde{\phi}_a(p)\big(p^2+\mu^2\big)\tilde{\phi}_a(-p)&=&\Lambda^2\int_{\bar{p}\leq 1}\tilde{\phi}_a(\Lambda \bar{p})\big(\bar{p}^2+\bar{\mu}^2\big)\tilde{\phi}_a(-\Lambda \bar{p}).
%\end{eqnarray}

\paragraph{The Quartic Action:}

Next we study the quartic term. We will assume from the start that $u_0=v_0$. By applying rules $1$, $2$ and $3$ of Wilson contraction to the $4-$point function, we get after a long calculation the approximation

\begin{eqnarray}
\Delta \tilde{V}(p_1,p_2,p_3,p_4)&=&4u_0 \cos\frac{p_1\wedge p_2}{2}\cos\frac{p_3\wedge p_4}{2}\times \Lambda^{d-4} V_d\Delta^2(\bar{\lambda})\bigg[1-3\bar{u}_0V_d\Delta^2(\bar{\lambda})~\bigg(1+\tilde{X}_{d-1}(\bar{\theta} \bar{\lambda}^2)\bigg)\nonumber\\
&+&10\bar{u}_0^2V_d^2\Delta^4(\bar{\lambda})~\bigg(1+\tilde{X}_{d-1}(\bar{\theta} \bar{\lambda}^2)\bigg)^2-35\bar{u}_0^3V_d^3\Delta^6(\bar{\lambda})~\bigg(1+\tilde{X}_{d-1}(\bar{\theta} \bar{\lambda}^2)\bigg)^3+...\bigg].\nonumber\\
\end{eqnarray}
%\begin{eqnarray}
%C_4(x)=\frac{1-(1+4x)^{-\frac{1}{2}}}{2x}=1-3x+10x^2-35x^3+...
%\end{eqnarray}
In this case, the expansion between brackets is recognized as the expansion of the connected $4-$point function of the zero-dimensional vector model which is given by the function \cite{Hikami:1978ya,Nishigaki:1996ts}
\begin{eqnarray}
C_4(a)=\frac{1-(1+4a)^{-\frac{1}{2}}}{2a}=1-3a+10a^2-35a^3+...
\end{eqnarray}
The effective coupling constant $a$ is still given by equation (\ref{a}).
%\begin{eqnarray}
%a=\bar{u}_0v_d(\rho)\Delta^2(\bar{\lambda})~\bigg(1+\tilde{X}_{d-1}(\bar{\theta} \bar{\lambda}^2)\bigg).
%\end{eqnarray}
The final result is therefore of the form
\begin{eqnarray}
\Delta \tilde{V}(p_1,p_2,p_3,p_4)&=&4\bar{u}_0 v_d(\rho)\Delta^2(\bar{\lambda}) C_4(a)\cos\frac{p_1\wedge p_2}{2}\cos\frac{p_3\wedge p_4}{2}.
\end{eqnarray}
The effective quartic action becomes
\begin{eqnarray}
S_{\rm intera}%&=&\frac{u_0}{N}\int_{p_1\leq \rho\Lambda}...\int_{p_4\leq \rho\Lambda }\tilde{\phi}_a(p_1)\tilde{\phi}_a(p_2)\tilde{\phi}_b(p_3)\tilde{\phi}_b(p_4)~(2\pi)^d\delta^d(p_1+...+p_4)~\tilde{V}_{\rm eff}(p_1,..,p_4)\nonumber\\
&=&\frac{\bar{u}_0}{N}\Lambda^{2d+4}\int_{p_1\leq \rho}...\int_{p_4\leq \rho}\tilde{\phi}_a(\Lambda p_1)\tilde{\phi}_a(\Lambda p_2)\tilde{\phi}_b(\Lambda p_3)\tilde{\phi}_b(\Lambda p_4)~(2\pi)^d\delta^d(p_1+...+p_4)~\tilde{V}_{\rm eff}(\Lambda p_1,..,\Lambda p_4).\nonumber\\\label{quarticS}
\end{eqnarray}
\begin{eqnarray}
\tilde{V}_{\rm eff}(\Lambda p_1,..,\Lambda p_4)&=&\bigg(1-4\bar{u}_0 V_d\Delta^2(\bar{\lambda}) C_4(a)\bigg)\cos\Lambda^2 \frac{p_1\wedge p_2}{2}\cos\Lambda^2 \frac{p_3\wedge p_4}{2}\nonumber\\
&+&\frac{1}{2}\bigg(\cos\Lambda^2\frac{p_1\wedge p_3+p_2\wedge p_4}{2}+\cos\Lambda^2\frac{p_1\wedge p_4+p_2\wedge p_3}{2}\bigg).
\end{eqnarray}
As before we will choose $\bar{\lambda}=1$. Notice that there is no correction to the non-planar vertex.
\paragraph{The Renormalization Group Equations:}
By putting (\ref{quadraticS}) and (\ref{quarticS}) together and making the rescaling $p\longrightarrow \rho p$ so that the range of the momentum returns to the original interval $[0,\Lambda]$ we obtain (with $\Lambda_{\rho}=\rho\Lambda$)

\begin{eqnarray}
S_{\rm eff}&=&\Lambda_{\rho}^{d+2}\int_{{p}\leq 1}\tilde{\phi}_a(\Lambda_{\rho} {p})\bigg[{p}^2+\hat{\mu}^2\bigg]\tilde{\phi}_a(-\Lambda_{\rho} {p})+\Lambda_{\rho}^{2d+4}\int_{p_1\leq 1}...\int_{p_4\leq 1}\tilde{\phi}_a(\Lambda_{\rho} p_1)\nonumber\\
&\times &\tilde{\phi}_a(\Lambda_{\rho} p_2)\tilde{\phi}_b(\Lambda_{\rho} p_3)\tilde{\phi}_b(\Lambda_{\rho} p_4)~(2\pi)^d\delta^d(p_1+...+p_4)~\hat{V}_{\rm eff}(\Lambda_{\rho} p_1,..,\Lambda_{\rho} p_4).
\end{eqnarray}
The renormalized mass and vertex are given by $\hat{\mu}^2=\rho^{-2}(\bar{\mu}^2+\delta\bar{\mu}^2)$ and $\hat{V}_{\rm eff}=\rho^{d-4}\tilde{V}_{\rm eff}$.
%\begin{eqnarray}
%\hat{\mu}^2&=&\rho^{-2}(\bar{\mu}^2+\delta\bar{\mu}^2).
%\end{eqnarray}
%\begin{eqnarray}
%\hat{V}_{\rm eff}(\Lambda_{\rho} p_1,..,\Lambda_{\rho} p_4)=\rho^{d-4}\tilde{V}_{\rm eff}(\Lambda_{\rho} p_1,..,\Lambda_{\rho} p_4).
%\end{eqnarray}
%This last equation is equivalent to the two equations
%\begin{eqnarray}
%\hat{u}_0=\rho^{d-4}\bar{u}_0\bigg(1-4\bar{u}_0 V_d\Delta^2(1) C_4(a)\bigg).
%\end{eqnarray}
%\begin{eqnarray}
%\hat{v}_0=\rho^{d-4}\bar{u}_0.
%\end{eqnarray}
We have then the renormalization group equations (with $\epsilon=4-d$)
\begin{eqnarray}
\hat{\mu}^2&=&\rho^{-2}\bigg(\bar{\mu}^2+(1+\bar{\mu}^2)\frac{(1+4a)^{\frac{1}{2}}-1}{1+\tilde{X}_{d-1}(\bar{\theta})}\bigg).
\end{eqnarray}
\begin{eqnarray}
\hat{u}_0&=&\rho^{-\epsilon}\bar{u}_0\bigg(1-2\frac{1-(1+4a)^{-\frac{1}{2}}}{1+\tilde{X}_{d-1}(\bar{\theta})}\bigg).
\end{eqnarray}
\begin{eqnarray}
\hat{v}_0=\rho^{-\epsilon}\bar{u}_0.
\end{eqnarray}
This result is supposed to be exact in the sense that all leading diagrams in the large $N$ limit were included in deriving it. However our estimation of these diagrams uses approximations which may or may not be appropriate. It was shown in  \cite{Hikami:1978ya,Nishigaki:1996ts} that in the commutative theory the same set of approximations reproduces the correct answer for vector models in the limit $\rho \longrightarrow 0$. The main point we want to stress here is that in the calculation which led to the above  recursion equations we did not involve any $\theta\longrightarrow 0$ limit, although we have argued that rule $1$ of Wilson contraction will be valid only in this limit. 

Two more remarks are in order. First note that the correct coupling constant in the commutative theory is actually $\bar{u}_0+\bar{v}_0$ and recall that we have chosen $\bar{v}_0=\bar{u}_0$. Thus it is natural to work with the combinations $\bar{u}_0\pm \bar{v}_0$ instead of $\bar{u}_0$ and $\bar{v}_0$. 
%In terms of the combinations $\bar{u}_0\pm \bar{v}_0$ the  renormalization group equations become
%\begin{eqnarray}
%\hat{u}_0+\hat{u}_0^{'}&=&\frac{2\rho^{-\epsilon}\bar{u}_0}{(1-\delta Z)^2}\bigg(1-\frac{1-(1+4a)^{-\frac{1}{2}}}{1+\tilde{X}_{d-1}(\bar{\theta})}\bigg).
%\end{eqnarray}
%\begin{eqnarray}
%\hat{u}_0-\hat{u}_0^{'}=-\frac{2\rho^{-\epsilon}\bar{u}_0}{(1-\delta Z)^2}\frac{1-(1+4a)^{-\frac{1}{2}}}{1+\tilde{X}_{d-1}(\bar{\theta})}.
%\end{eqnarray}
%The second remark concerns the wave function renormalization $\delta Z$. We recall that the set of approximations (the so-called Wilson contraction) used in estimating the leading diagrams in the large $N$ limit consists in three rules where the first rule says that any external momentum which is wedge with an internal momentum must be set to zero. This rule was, in fact, not used in the estimation of the tadpole graphs contributing to the $2$-point function and as a consequence the calculation of the $2$-point function is, in a sense, more exact. Using this rule in the $2$-point function would simply  imply that we must set $\delta Z=1$.

The second remark concerns the role of the non-commutativity parameter $\bar{\theta}$. In this article we will think of the non-commutativity parameter $\bar{\theta}$ in the same way as the dimension $d$. The fixed point and the critical exponents will therefore depend on the values of $\bar{\theta}$ and $d$, i.e. these parameters do not receive renormalization. There is also the alternative picture (which we will not pursue here) to consider $\bar{\theta}$ on equal footing with the other coupling constants of the theory. 
%We note also here that there is also the  parameter $\rho$ which we must dispose of it at the end of the calculation.

We rewrite the renormalization group equations in the form
%\begin{eqnarray}
%\hat{\mu}^2&=&\rho^{-2}\bigg(\bar{\mu}^2+(1+\bar{\mu}^2)\frac{(1+4a)^{\frac{1}{2}}-1}{1+\tilde{X}_{d-1}(\bar{\theta})}\bigg).
%\end{eqnarray}
\begin{eqnarray}
\hat{\mu}^2&=&\rho^{-2}\bigg(\bar{\mu}^2+t(1+\bar{\mu}^2)\frac{(1+4a)^{\frac{1}{2}}-1}{2}\bigg).
\end{eqnarray}
%\begin{eqnarray}
%\hat{u}_0+\hat{u}_0^{'}&=&2\rho^{-\epsilon}\bar{u}_0\bigg(1-\frac{1-(1+4a)^{-\frac{1}{2}}}{1+\tilde{X}_{d-1}(\bar{\theta})}\bigg).
%\end{eqnarray}
\begin{eqnarray}
\hat{u}_0+\hat{v}_0&=&2\rho^{-\epsilon}\bar{u}_0\bigg(1-t\frac{1-(1+4a)^{-\frac{1}{2}}}{2}\bigg).
\end{eqnarray}
%\begin{eqnarray}
%\hat{u}_0-\hat{u}_0^{'}=-2\rho^{-\epsilon}\bar{u}_0\frac{1-(1+4a)^{-\frac{1}{2}}}{1+\tilde{X}_{d-1}(\bar{\theta})}.
%\end{eqnarray}
\begin{eqnarray}
\hat{u}_0-\hat{v}_0=-\rho^{-\epsilon}\bar{u}_0t(1-(1+4a)^{-\frac{1}{2}}).
\end{eqnarray}
We have set
\begin{eqnarray}
t=\frac{2}{1+\tilde{X}_{d-1}(\bar{\theta})}=1+\frac{\bar{\theta}^2}{4d}+\frac{\bar{\theta}^4}{8d^2(d+2)}+....
\end{eqnarray}
We will also use the notation
\begin{eqnarray}
T=t\frac{(1+4a)^{\frac{1}{2}}-1}{2}.
\end{eqnarray}
 \paragraph{The Commutative Fixed Point:}
The commutative theory is characterized by $\bar{\theta}=0$ or equivalently $t=1$.  A non-gaussian fixed point is given by $\hat{\mu}^2=\bar{\mu}^2=\mu_*^2$ and $\hat{u}_0+\hat{v}_0=2\bar{u}_0=2u_*$ or equivalently
\begin{eqnarray}
\mu_*^2=\frac{T_*}{\rho^2-1-T_*}.
\end{eqnarray}

\begin{eqnarray}
\rho^{\epsilon}&=&1-\frac{1-(1+4a_*)^{-\frac{1}{2}}}{2}.\label{contr1c}
\end{eqnarray}
In the limit $\epsilon\longrightarrow 0$, we have a solution for any value of the dilatation parameter $\rho$ given by
\begin{eqnarray}
a_*=-\epsilon\ln\rho.
\end{eqnarray}
The fixed point near $\epsilon\simeq 0$ is
\begin{eqnarray}
\mu_*^2=-\frac{\epsilon\ln\rho}{\rho^2-1}~,~u_*=-\frac{\epsilon\ln\rho}{2V_d}.
\end{eqnarray}
The condition (\ref{contr1c}), for $0<\epsilon\leq 2$, leads to the solution
\begin{eqnarray}
 a_*=\frac{(1-\rho^{\epsilon})\rho^{\epsilon}}{(1-2\rho^{\epsilon})^2}.
\end{eqnarray}
Therefore the fixed point is given by (with $u_*=(1+\mu_*^2)^2a_*/2V_d$)
\begin{eqnarray}
\mu_*^2&=&\frac{1-\rho^{\epsilon}}{-(\rho^2+\rho^{\epsilon})+2\rho^{\epsilon+2}}~,~
u_*=\frac{1}{2V_d}\frac{(1-\rho^2)^2(1-\rho^{\epsilon})\rho^{\epsilon}}{\big(-(\rho^2+\rho^{\epsilon})+2\rho^{\epsilon+2}\big)^2}.
\end{eqnarray}
By inspection we must have $\rho^{\epsilon}\geq 1/2$ and $\rho^{\epsilon}\leq 1$ for $T_*$ to be positive definite and as a consequence $\mu_*^2\leq 0$. Clearly we have always $u_*\geq 0$.

The main observation of \cite{Ferretti:1995zn} is that we can/must extrapolate the above fixed point to the second sheet $\rho\in[0,1/2]$ in order to reach  the non-perturbative Wilson-Fisher fixed point and recover all critical exponents of large vector models in dimensions $2\leq d \leq 4$ in the limit $\rho\longrightarrow 0$. This limit corresponds to integrating all modes of the theory.  As pointed out in \cite{Ferretti:1995zn}, the theory flows into the infrared limit very quickly so to overcome the errors of the approximation when the cutoff theory is very strongly coarse-grained by  taking one single step of the renormalization group transformation.

On the perturbative sheet the function $a_*=f(z)$ with $z=\rho^{\epsilon}$ starts from $f=0$ at $z=1$ then increases to $\infty$ as $z$ decreases to $z=1/2$. Clearly we can not take the limit $\rho\longrightarrow 0$ unless we continue the fixed point to the second (non-perturbative) sheet. On the non-perturbative sheet the function $f(z)$ starts from $f=0$ at $z=0$ then increases to $\infty$ as $z$ increases to $z=1/2$.  We compute
\begin{eqnarray}
\sqrt{1+4a_*}=\pm\frac{1}{2\rho^{\epsilon}-1}.
\end{eqnarray}
The plus sign corresponds to the perturbative solution (\ref{contr1c}). The minus sign leads to
\begin{eqnarray}
1-\rho^{\epsilon}&=&1-\frac{1-(1+4a_*)^{-\frac{1}{2}}}{2}.\label{contr2c}
\end{eqnarray}
In this case  $\rho^{\epsilon}$ is in the range $[0,1/2]$. As a consequence we can continue the solution from the first sheet to the second sheet where $a_*$ goes from $0$ at $\rho^{\epsilon}=1$ to $\infty$ at $\rho^{\epsilon}=1/2$ and then back to $0$ at $\rho^{\epsilon}=0$. On the second sheet, we have $T_*=\rho^{\epsilon}/(1-2\rho^{\epsilon})$ and as a consequence the critical values $\mu_*^2$ and $u_*$ are given by
\begin{eqnarray}
\mu_*^2&=&-\frac{\rho^{\epsilon}}{1-\rho^2-\rho^{\epsilon}+2\rho^{2+\epsilon}}<0~,~
u_*=\frac{1}{2V_d}\frac{\rho^{\epsilon}(1-\rho^{\epsilon})(1-\rho^2)^2}{(1-\rho^2-\rho^{\epsilon}+2\rho^{2+\epsilon})^2}>0.
\end{eqnarray}
This is the commutative Wilson-Fisher fixed point. The next step is to linearize the renormalization group equations around this fixed point and derive the critical exponents as a function of $\rho$. In the limit $\rho\longrightarrow 0$ we recover the correct behavior. This has been done in some detail in \cite{Ferretti:1995zn}.
\paragraph{The Non-Commutative Fixed Point:}

Again,  a non-gaussian fixed point is given by $\hat{\mu}^2=\bar{\mu}^2=\mu_*^2$ and $\hat{u}_0+\hat{v}_0=2\bar{u}_0=2u_*$ or equivalently
\begin{eqnarray}
\mu_*^2=\frac{T_*}{\rho^2-1-T_*}.
\end{eqnarray}

\begin{eqnarray}
\rho^{\epsilon}&=&1-t\frac{1-(1+4a_*)^{-\frac{1}{2}}}{2}.\label{contr1}
\end{eqnarray}
In the limit $\epsilon\longrightarrow 0$, we have a solution for any value of the dilatation parameter $\rho$ given by
\begin{eqnarray}
a_*=-\epsilon\ln\rho.
\end{eqnarray}
The fixed point near $\epsilon\simeq 0$ is
\begin{eqnarray}
a_*=-\frac{\epsilon\ln\rho}{t}~,~\mu_*^2=-\frac{\epsilon\ln\rho}{\rho^2-1}~,~u_*=-\frac{\epsilon\ln\rho}{2tV_d}.
\end{eqnarray}
For  $0<\epsilon\leq 2$, the critical value of $a$ is given by
\begin{eqnarray}
 a_*=f(z)=\frac{(1-z)(t-1+z)}{(t-2+2z)^2}~,~z=\rho^{\epsilon}.\label{eqnt}
\end{eqnarray}
On the perturbative sheet the function $f(z)$ starts from $f=0$ at $z=1$ then increases to $\infty$ as $z$ decreases to $z=1-{t}/{2}$. Clearly we can not take the limit $\rho\longrightarrow 0$ unless we continue the fixed point to the second (non-perturbative) sheet. On the non-perturbative sheet the function $f(z)$ starts from $f=(t-1)/(t-2)^2>0$ at $z=0$ then increases to $\infty$ as $z$ increases to $z=1-{t}/{2}$. The function $f(z)$ vansihes at $z=1-t<0$. We compute
\begin{eqnarray}
\sqrt{1+4a_*}=\pm\frac{t}{t-2+2z}.\label{sheets}
\end{eqnarray}
The plus sign corresponds to the perturbative solution (\ref{contr1}). We have on the perturbative sheet
\begin{eqnarray}
T_*=t\frac{1-\rho^{\epsilon}}{t-2+2\rho^{\epsilon}}.
\end{eqnarray}
We must clearly have $\rho^{\epsilon}\geq 1-t/2$ and $\rho^{\epsilon}\leq 1$ for $T_*$ to be positive definite. We get therefore
\begin{eqnarray}
\mu_*^2&=&t\frac{1-\rho^{\epsilon}}{2(1-t)+(t-2)(\rho^2+\rho^{\epsilon})+2\rho^{\epsilon+2}}~,~\nonumber\\
u_*&=&\frac{t}{2V_d}\frac{(1-\rho^2)^2(1-\rho^{\epsilon})(t-1+\rho^{\epsilon})}{\big(2(1-t)+(t-2)(\rho^2+\rho^{\epsilon})+2\rho^{\epsilon+2}\big)^2}. \label{fixedP}
\end{eqnarray}
We can check that we have always  $u_*\geq 0$ and  $\mu_*^2\leq 0$.
%Clearly we have always $u_*\geq 0$. However, the sign of $\mu_*^2$ in $d>2$ is always found to be negative  while at $d=2$ it is negative as long as $t<2$ or equivalently $\bar{\theta}<\pi$. We will restrict ourselves therefore to the range $1\leq t\leq 2$ or equivalently $0\leq \bar{\theta}\leq \pi$.

 The minus sign in (\ref{sheets}) leads to
\begin{eqnarray}
2-t-\rho^{\epsilon}&=&1-t\frac{1-(1+4a_*)^{-\frac{1}{2}}}{2}.\label{contr2}
\end{eqnarray}
By setting $z=\rho^{\epsilon}$ on the first sheet and $z=2-t-\rho^{\epsilon}$ on the second sheet, equations (\ref{contr1}) and (\ref{contr2}) become the equation 
\begin{eqnarray}
z&=&1-t\frac{1-(1+4a_*)^{-\frac{1}{2}}}{2}.
\end{eqnarray}
This leads to $ a_*=f(z)$ where $f(z)$ is given by equation (\ref{eqnt}). The first sheet corresponds to the interval $z\in[1-t/2,1]$ whereas the second sheet corresponds to $z\in[0,1-t/2]$. In this case when we continue the solution to the second sheet we observe that the critical coupling constant $a_*$ does not return to $0$ when we take the limit $\rho^{\epsilon}\longrightarrow 0$. Indeed $f(0)=(t-1)/(t-2)^2>0$. As long as $\theta$ is sufficiently small we have $t$ near $1$ and as a consequence $f(0)$ is small and the argument of the commutative theory essentially goes through. 

The value of the critical value $T_*$ on the second sheet is
 \begin{eqnarray}
T_*&=&-\frac{t(t-1+\rho^{\epsilon})}{t-2+2\rho^{\epsilon}}.
\end{eqnarray}
As a consequence the critical values $\mu_*^2$ and $u_*$ on the second sheet are given by
\begin{eqnarray}
\mu_*^2&=&-\frac{t(t-1+\rho^{\epsilon})}{2-2t+t^2+(t-2)(\rho^2+\rho^{\epsilon})+2\rho^{2+\epsilon}}~,~\nonumber\\
u_*&=&\frac{t}{2V_d}\frac{(t-1+\rho^{\epsilon})(1-\rho^{\epsilon})(1-\rho^2)^2}{\big(2-2t+t^2+(t-2)(\rho^2+\rho^{\epsilon})+2\rho^{2+\epsilon}\big)^2}.
\end{eqnarray}
This is the non-commutative Wilson-Fisher fixed point. 

From the other hand we observe that for a fixed $t$ the limit of the perturbative fixed point when $\rho^{\epsilon}\longrightarrow 1$ is (with $1/\tilde{V}_d=2^{d-1}\pi^{d/2}\Gamma(d/2)$)
\begin{eqnarray}
a_*=-\frac{\epsilon\ln\rho}{t}~,~\mu_*^2=-\frac{\epsilon}{2}~,~u_*=\frac{\epsilon}{2\tilde{V}_d}.
\end{eqnarray}
The limit for a fixed $t$ of the non-commutative Wilson-Fisher fixed point when $\rho^{\epsilon}\longrightarrow 0$ is
\begin{eqnarray}
a_*&=&\frac{t-1}{(t-2)^2}[1+\frac{\rho^{\epsilon}}{t-1}-\rho^{\epsilon}-\frac{4}{t-2}\rho^{\epsilon}+...]~,~\nonumber\\
\mu_*^2&=&-\frac{t(t-1)}{(t-1)^2+1}[1+\frac{\rho^{\epsilon}}{t-1}-\frac{t-2}{(t-1)^2+1}\rho^{\epsilon}+...]~,~\nonumber\\
u_*&=&\frac{1}{2V_d}\frac{t(t-1)}{((t-1)^2+1)^2}[1+\frac{\rho^{\epsilon}}{t-1}-\rho^{\epsilon}-2\frac{t-2}{(t-1)^2+1}\rho^{\epsilon}+...].
\end{eqnarray}
In contrast with the commutative theory and with the perturbative fixed point, the non-commutative Wilson-Fisher fixed point is not vanishingly small in the limit $\rho^{\epsilon}\longrightarrow 0$ and becomes significantly more important as we increase $t$ from $1$ to $2$, i.e. as we increase the non-commutativity $\bar{\theta}$ from $0$ to $\pi$. Indeed, we see that $a_*\longrightarrow \infty$ when $t\longrightarrow 2$, i.e. when we have only one sheet $[0,1]$. Putting it differently, in the limit $t\longrightarrow 2$ the two-sheeted structure of $a_*$ disappears and we end up only with the perturbative fixed point (\ref{fixedP}). 

% and as a consequence the fixed points at $\rho^{\epsilon}=0$ and $\rho^{\epsilon}=1$ seem to be different. 

\section{Conclusion and Outlook}
In this article we have studied the non-commutative vector sigma model with the most general $\phi^4$ interaction on the Moyal-Weyl spaces ${\bf R}_{\theta}^d$ in the large $N$ limit. We computed the $2-$ and $4-$point functions to all orders and then applied the approximate Wilson renormalization group recursion formula to study the renormalized coupling constants of the theory. More precisely, we have taken into account all leading Feynman diagrams in the large $N$ limit, and then estimated them using the so-called Wilson contraction which consists of three rules. We have argued that the first rule of Wilson contraction is certainly valid in the limit $\bar{\theta}\longrightarrow 0$.

In the commutative theory we observe that the non-perturbative Wilson-Fisher fixed point is located on the second sheet of the parameter $a$ which is the coupling constant of the zero dimensional reduction of the theory. The Wilson-Fisher fixed point can be reached in the limit in which we send the dilation parameter $\rho$ to zero which corresponds to a single step of the renormalization group transformation. The Wilson-Fisher fixed point is also found to scale to zero in the limit $\rho\longrightarrow 0$ although differently then the perturbative fixed point in the limit $\rho\longrightarrow 1$.

In the non-commutative theory the Wilson-Fisher fixed point does not scale to zero  in the limit $\rho\longrightarrow 0$ in contrast with the perturbative fixed point which still scales to zero as $\rho\longrightarrow 1$. The main obstacle comes from the fact that the critical coupling constant $a_*$ which starts from $0$ at $\rho^{\epsilon}=1$, increases to $\infty$ at $\rho^{\epsilon}=1-t/2$, does not return to zero as we decrease $\rho^{\epsilon}$ back from $\rho^{\epsilon}=1-t/2$ to $0$. 

The non-perturbative window (sheet) shrinks as we increase the non-commutativity until it disappears at $t=2$. At this point, the critical value $a_*$ diverges and the non-commutative Wilson-Fisher point becomes very different. It is natural to conjecture that the point at $t=2$ is completely different (lies in a different universality class) then the commutative Wilson-Fisher fixed point at $t=1$ which is in the Ising universality class. Therefore the non-commutative Wilson-Fisher fixed point interpolates between these two classes.

A thorough study of the renormalized action (including the coupling constant $v$) and the non-commutative Wilson-Fisher fixed  point (including the range  $t>2$), together with the computation of the critical exponents, and improvement  of the approximate Wilson renormalization group recursion formula will be reported elsewhere \cite{inprogress}. For example, already we observe from our study here that we can compute the $2-$point function without any resort to the first rule of Wilson contraction resulting in equation (\ref{wave}) which indicates the presence of a wave function renormalization in the theory. As already discussed in \cite{Chen:2001an} this leads to an instability of the theory. Solving the renormalization group equations in the case of a non-zero wave function may/will require a numerical approach. 

Another approach to the Wilson renormalization group recursion formula of non-commutative $\phi^4$ will be in terms of the matrix model associated with the  theory which will allow us to access the limit $\bar{\theta}\longrightarrow \infty$ instead. We also hope to return to this point in  future communication \cite{inprogress}.

\paragraph{Acknowledgments:}
This research was supported by ``The National Agency for the Development of University Research (ANDRU)'' under PNR contract number  U23/Av58 (8/u23/2723).

\end{document}